\documentclass[%
 reprint,
 amsmath,amssymb,
 aps,
]{revtex4-2}

\usepackage{graphicx}% Include figure files
\usepackage{dcolumn}% Align table columns on decimal point
\usepackage{bm}% bold math
\begin{document}

\preprint{APS/123-QED}

\title{Enhancing Precision of Signal Correction in PVES Experiments: The Impact of Bayesian Analysis on the Results of the QWeak and MOLLER Experiments}% Force line breaks with \\
\author{E. Gorgannejad}
%\collaboration{MOLLER Collaboration}
%\altaffiliation[Also at ]{Physics Department, University of Manitoba.}
\author{W. Deconinck}
%\email{wouter.deconinck@umanitoba.ca}
\affiliation{University of Manitoba, Winnipeg, MB R3T2N2 Canada}%
%\collaboration{QWeak and MOLLER Collaborations}
\author{D.S. Armstrong}
%\homepage{http://www.Second.institution.edu/~Charlie.Author.}
\affiliation{William \& Mary, Williamsburg, Virginia 23185, USA}
%\collaboration{QWeak and MOLLER Collaborations}%\noaffiliation

\date{\today}% It is always \today, today,
\begin{abstract}
The precise measurement of parity-violating asymmetries in parity-violating electron scattering experiments is a powerful tool for probing new physics beyond the Standard Model. Achieving the expected precision requires both experimental and post-processing signal corrections. This includes using auxiliary detectors to distinguish the main signal from background signals and implementing post-measurement corrections, such as the Bayesian statistics method, to address uncontrolled factors during the experiments. Asymmetry values in the scattering of electrons off proton targets in QWeak and P2 and off electron targets in MOLLER are influenced by detector array configurations, beam polarization angles, and beam spin variations. The Bayesian framework refines full probabilistic models to account for all necessary factors, thereby extracting asymmetry values and the underlying physics under specified conditions. For the QWeak experiment, a reanalysis of the inelastic asymmetry measurement using the Bayesian method has yielded a closer fit to measured asymmetries, with uncertainties reduced by 40\% compared to the Monte Carlo minimization method. This approach was successfully applied to simulated data for the MOLLER experiment and is predicted to be similarly effective in P2.
%\begin{description}
%\end{description}
\end{abstract}
%Use showkeys class option if keyword
\maketitle
%\tableofcontents
\section{\label{Introduction}Introduction}
Advancements in polarized electron accelerators and measurement techniques have enabled low-energy electron scattering experiments involving longitudinally polarized electrons colliding with unpolarized targets to access both electromagnetic and weak interactions. Unlike electromagnetic interactions, weak interactions do not preserve parity, leading to measurable parity violations. These violations allow the investigation of  Standard Model (SM) parameters, particularly the Z boson couplings \cite{kumar2013low}. Additionally, they provide accurate studies of nuclear properties \cite{erler2014weak} and contribute to the search for new parity-violating interactions beyond the SM \cite{ramsey1999low}.

Historically, the first observation of parity violation in electron scattering was through the pioneering E122 experiment conducted at the Stanford Linear Accelerator Centre (SLAC) \cite{prescott1978parity}. This marked the establishment of the Parity Violating Electron Scattering (PVES) experiment field. These experiments aim to measure spin-dependent scattering asymmetries when the polarization spin-axis (helicity) of electrons is flipped relative to their momentum at the target. The asymmetries associated with parity-violating phenomena are typically on the order of several parts per million (ppm) or even smaller, necessitating precise control of the electron beam. Technological advances have been crucial in facilitating precise helicity reversals without altering other beam characteristics, including current, position, energy, and size.

The Continuous Electron Beam Accelerator Facility (CEBAF) \cite{adderley2023overview} employs superconducting linacs to deliver a continuous electron beam, resulting in stable beam conditions ideal for conducting parity-violation experiments. CEBAF has hosted significant experiments such as the Qweak experiment \cite{carlini2012qweak}, which measured the weak charge of the proton using a 1.165 GeV electron beam at 180 µA, achieving an asymmetry measurement precision of \( A_{PV} = -226.5 \pm 9.3 \) ppb (parts per billion). Moreover, the upcoming MOLLER experiment \cite{benesch2014moller}, planned to begin in 2026, will utilize an 11 GeV electron beam with a current of 65 µA to probe the weak charge of the electron, aiming to measure an expected parity violation of \(A_{PV} \approx -32\) ppb with a precision of \( \pm 0.54 \) ppb. Additionally, the Mainz Energy-Recovering Superconducting Accelerator (MESA) \cite{hug2020mesa} will host the P2 experiment \cite{Becker_2018} through its new experimental hall and innovative beam manipulation techniques. This setup will enable P2 to explore parity violations using a beam of 150 $\mu$A at an energy of 155 MeV, targeting the measurement of the expected parity-violating asymmetry of \( A_{PV} \approx -40 \) to a precision of \( \pm 0.56 \) ppb, the goal being an improved determination of the proton's weak charge. 

In conducting these precision experiments, integration mode, where signals are accumulated over a periodic time window, is predominantly used to measure the scattered electrons at an unprecedentedly high rate on the order of 10's of GHz. The measured asymmetry value necessitates corrections before it can be interpreted as a measure of parity violation in electron scattering. The corrections accommodate background effects from other scattering processes, specifically accounting for dilution factors and asymmetries. These parameters contribute to the correction of the parity-violating asymmetry, $A_{PV}$, using 
\begin{equation}
A_{PV} = \frac{\frac{A_{\rm expt}}{P_b} - \sum_i f_i^{\rm bkgd} A_i^{\rm bkgd}}{1-\sum_i f_i^{\rm bkgd}}, 
\label{SignalCorrectionFormula}
\end{equation}
where \( A_{\rm expt} \) is the experimentally measured asymmetry, which must be corrected for background processes characterized by fractional dilution factors \( f_i^{\rm bkgd} \) and asymmetries \( A_i^{\rm bkgd} \), and \( P_b \) is the beam polarization.

In contrast to the integration mode, occasional data-taking in the counting mode records individual scattering events, providing detailed information on each occurrence. This allows precise background subtraction and the identification of anomalous events that could impact the primary measurement. 

The precision level of signal corrections in PVES experiments necessitates not only technological advancements but also methodological innovation. The transition from traditional frequentist methods \cite{casella2024statistical}, which interpret probability as the long-term frequency of events, to a Bayesian framework \cite{chivers2024everything} represents a significant change. In Bayesian analysis, probability is understood as a degree of belief regarding the occurrence of an event or the validity of a hypothesis. In this paper we show that this shift is important in high-precision experiments like QWeak \cite{carlini2012qweak}, MOLLER \cite{benesch2014moller}, and P2 \cite{Becker_2018}, where the measurement of asymmetries requires a refined approach to uncertainty. 

The basis of Bayesian statistics was first described in 1763 by Reverend Thomas Bayes and published by Richard Price on inverse probability \cite{bayes1763lii}. In 1825, Pierre Simon Laplace published the Bayes’ theorem \cite{laplace1814essai}. In the past 50 years, the ideas of inverse probability and Bayes’ theorem have been prominent tools in applied statistics although they are long-standing in mathematics. Bayesian methods of data analysis are now widely used across different fields of science such as ecology\cite{ellison2004bayesian}, social and behavioural sciences\cite{van2017systematic}, genetics\cite{stephens2009bayesian}, medicine\cite{ashby2006bayesian}, educational research\cite{thurlings2017learning}, epidemiology\cite{rietbergen2017reporting}, organizational sciences\cite{kruschke2014time},\cite{smid2020bayesian}, modeling\cite{rupp2004bayes}, nuclear physics\cite{yang2020bayesian}, experimental data analysis\cite{lecoutre2011bayesian}, and the experimental particle physics\cite{golchi2018frequency}. However, Bayesian methods have not yet been applied in PVES experiments. 

Bayesian statistics utilize Bayes' theorem to integrate prior information with newly observed data into a posterior distribution. This approach enables direct probabilistic inferences and enhances the accuracy of signal corrections. It accounts for correlations between parameters, such as asymmetry components, by forming the posterior based on measured data, thereby providing a more accurate estimation of these parameters. This analysis method offers a robust framework for making informed inferences, contributing to a deeper understanding of the underlying phenomena in PVES experiments.

In this paper, we explore the application of Bayesian statistical methods to improve the precision of signal corrections in the QWeak and MOLLER electron scattering experiments. Our methodology incorporates establishing prior distributions and constructing likelihood functions, followed by using Markov Chain Monte Carlo (MCMC) techniques. We detail the reanalysis of an ancillary measurement to the QWeak experiment and the simulation-based preparations for the MOLLER experiment, demonstrating the enhanced accuracy provided by Bayesian approaches. Additionally, we employ covariance and correlation analysis to elucidate the interdependencies among experimental parameters, further refining our understanding of the underlying physical phenomena. 
\section{\label{Methodology}Methodology}
In this section, we explore the stages of Bayesian analysis, from establishing prior distributions and constructing models to drawing inferences using Bayes' theorem, specifically highlighting how this approach is implemented in data analysis of the QWeak and MOLLER experiments.
\subsection{\label{Bayesian Analysis}Bayesian Analysis}
The Bayesian approach studies conditional probability. When two occurring events A and B are dependent or conditional, the basic conditional probability can be written as
\begin{equation}
P(B\cap A) = P(B|A) P(A) ,
\label{BasicConditionalProbability1}
\end{equation}
where ${P(B\cap A )}$ represents the probability of occurrence of both B and A, $P(B|A)$ is the probability of event B conditional on the occurrence of event A, and $P(A)$ is the probability of occurring event A. While, generically,  $P(B|A) \not = P(A|B)$, however $P(B\cap A) = P(A\cap B)$, so we can also write
\begin{equation}
{P(A\cap B )}=P(A|B)P(B)  
\label{BasicConditionalProbability2}
\end{equation} which can be combined with \ref{BasicConditionalProbability1}, to yield 
%can be rewritten as:

\begin{equation}
P(A|B)=\frac{P(B|A)P(A)}{P(B)}  .
\label{BayesRule}
\end{equation}
Equation \ref{BayesRule} is Bayes’ rule. By extending these principles to the situation of data and model parameters, A would be the observation of specific data set $y$ and B is the realization of the model with parameters $\theta$. So, Equation \ref{BayesRule} is written as follows:\\
\begin{equation}
P(\theta|y)=\frac{P(y|\theta)P(\theta)}{P(y)} 
\label{BayesRuleOriginal}
\end{equation}
$P(\theta|y)$ is a conditional probability of the model parameters $\theta$ on the observation of data set $y$ and represents the posterior distribution. $P(y|\theta)$ is the conditional probability of the data given the model parameters and represents the likelihood function. Finally, $P(\theta)$ represents the probability of model parameter values in the population, known as the prior distribution and $P(y)$ is a normalizing factor. So, the posterior distribution is proportional to the likelihood function and the prior distribution. As a conclusion, the three basic steps of the typical Bayesian workflow are as follows: 
\begin{enumerate}
\item Choosing the prior distribution, which is normally chosen before data collection, to represent available knowledge about a certain parameter in a statistical model.
\item Choosing the parameter information present in the observed data to determine the likelihood.
\item Combining the prior distribution and the likelihood function to form the posterior using Bayes’ theorem. 
\end{enumerate}
This theorem demonstrates how beliefs are updated in light of new data and details how uncertainties regarding model parameters are quantified.

To implement Bayesian analysis in the QWeak and MOLLER experiments, we quantify the prior distribution based on existing knowledge, use experimental data to form the likelihood function, and select a model to link these to the parameters of interest, the asymmetry components. Regarding the prior distribution, $P(\theta)$ in Equation \ref{BayesRuleOriginal}, a noninformative prior knowledge is assumed for both experiments, represented by a Gaussian distribution with a very large uncertainty. This allows the measurements to be the dominant influence on the posterior distributions. Concerning the input data, $y$ in Equation \ref{BayesRuleOriginal}, the measured asymmetry values are crucial. These values depend on the configuration of the experimental setup and the number of detectors, which are detailed in subsequent sections. Alongside the measured asymmetry values, other parameters, such as the beam polarization angle and spin variation, should also be considered as input data. Finally, the models that link the input data to the model parameters differ between the experimental setups.

The inherent complexities and computational challenges of working with these high-dimensional models necessitate the use of approximation techniques. These techniques aim to simplify reality, thereby enhancing our understanding of the system's components. Various computational challenges in Bayesian inference have historically favoured frequentist methods until advancements in computational techniques allowed more efficient Bayesian methods. Techniques such as Approximate Bayesian Computation (ABC) \cite{sisson2018handbook}, Integrated Nested Laplace Approximations (INLA) \cite{munoz2013estimation}, and Variational Bayesian methods \cite{blei2017variational} have been developed. These methods generally approximate only marginal posterior distributions of individual parameters, not the joint distributions of multiple parameters with their potential correlations. MCMC methods, introduced by Gelfand and Smith in 1990 \cite{gelfand1990sampling}, have become prominent for their ability to approximate the joint posterior distribution through sampling. In these methods, which are computationally efficient and widely used in Bayesian analysis, the Monte Carlo part denotes the sampling process, and the Markov Chain part describes the mechanism for obtaining these samples.

A variety of software packages implement MCMC methods. In this research, we utilize Stan with Python \cite{carpenter2017stan}, a robust Bayesian modelling language that includes the Hamiltonian Monte Carlo (HMC) and No U-Turn Sampler (NUTS) algorithms \cite{hoffman2014no}. Both provide efficient fits for complex models. HMC employs Hamiltonian dynamics to propose new sampling states, enabling large leaps across state spaces while maintaining high acceptance probabilities. This reduces autocorrelation in samples and improves efficiency, especially in high-dimensional spaces. However, HMC requires careful tuning of hyperparameters such as step size and the number of steps. NUTS enhances HMC by tuning these hyperparameters. 
\subsection{\label{QWeak Experiment}QWeak Experiment}
The QWeak experiment precisely measured the weak charge of the proton by measuring the parity-violating asymmetry in elastic electron-proton scattering at low momentum transfer. For an in-depth description of the QWeak experiment apparatus, see \cite{allison2015qweak}. In this setup, a polarized electron beam with an energy of 1.165 GeV and a current of 180 $\mu$A was aimed at a 35 cm liquid hydrogen target. The beam's polarization, about 85\%, was crucial for measuring the asymmetry in the scattering process. As electrons interacted with the target's protons, they could scatter elastically. The primary collimator selected scattered electrons for a particular range of scattering angles. Subsequently, the scattered electrons traversed a toroidal magnet which focused the elastically scattered electrons onto an array of eight fused-silica (quartz) main detectors (MDs), capturing the scattered electrons.

In a shorter ancillary measurement of the QWeak experiment \cite{androic2020parity}, the apparatus and experimental conditions were modified in two main ways compared to the weak-charge measurement: the beam energy was increased to 3.35 GeV to access the inelastic scattering kinematics of interest, and one of the main detectors was modified to enhance its sensitivity to pions. Due to the higher beam energy, a significant background was introduced in the main detectors, caused by negative pions produced in the target. Positively charged pions were swept out of the acceptance by the spectrometer’s magnetic field. %In the main QWeak experiment, 
Due to the high-rate integrating mode of the detector readout, it was not possible to separate the contributions of individual electrons and pions to the asymmetry measurement. Therefore, in the ancillary measurement, one of the main detectors was modified to enhance its sensitivity to pions in order to measure and correct for this pion background. The modification involved adding a 10.2 cm thick Pb absorber just upstream of the detector, which significantly attenuated the signal from scattered electrons without affecting the signal from the majority of pions. As a result, the asymmetry in MD7 was dominated by the incident pions, with a different mixture of electron and pion signals compared to the other seven main detectors.

As a result of the beam-delivery requirements for an experiment running concurrently in another experimental hall, this ancillary measurement utilized a beam with a polarization angle, \( P \), of \( -19.7^\circ \pm 1.9^\circ \) \cite{androic2020parity}, which was significantly deviated from the ideal longitudinal alignment. This introduced a large transverse component of about 33\%, influencing the physics asymmetry measurements. The setup with a beam that was neither purely longitudinal nor purely transverse in polarization, termed mixed data, involved 108 hours of data collection. For calibration, data-taking periods with essentially purely transverse polarization at \( 92.2^\circ \pm 1.9^\circ \) were conducted, referred to as transverse data, and comprised 4.3 hours of data-taking. Using eight main detectors (MDs) and two data-taking modes, the experiment measured sixteen asymmetry values, \( A_{i, j} \), as formulated below:
\begin{equation}
A_{i, j} = \frac{Y^+_{i, j} - Y^-_{i, j}}{Y^+_{i, j} + Y^-_{i, j}}
\label{SignalYields}
\end{equation}
where \( Y^{\pm} \) are the integrated photomultiplier (PMT) signal yields corresponding to right-handed/left-handed (\( \pm \)) helicity states. Here $i$ labels the detector and $j$ the data set (mixed or transverse). 
The measured asymmetries, \(A_{i, j}^{\text{meas}}\), accounting for pion contributions and background signals, were modelled by 
\begin{equation}
\begin{aligned}
A_{i, j}^{\text{meas}} &= (1 - f_{\text{NB}}^{i}) \Bigl[ (1 - f_{\pi}^i)(A_{e}^{L} \cos \theta_{P}^{j} +  A_{e}^{T} \sin \theta_{P}^{j} \sin \phi^{i}) \\ 
& + f_{\pi}^{i}(A_{\pi}^{L} \cos \theta_{P}^{j} + A_{\pi}^{T} \sin \theta_{P}^{j} \sin \phi^{i}) \Bigr].
\label{MeasuredAsymmetriesQweak}
\end{aligned}
\end{equation}
Here, \( f_{\pi}^i \) represents the fractional yield of pions at MD \( i \), and \( A_{e(\pi)}^{L} \) and \( A_{e(\pi)}^{T} \) denote the longitudinal and transverse asymmetries for electrons and pions, respectively. \( \theta_{P}^{j} \) is the beam polarization angle for the run type \( j \), and \( f_{\text{NB}}^{i} \) is the neutral background yield fraction for MD \( i \). The azimuthal angles, \( \phi^{i} \), define the placement of the MDs with specific values like \( \phi^{1} = 0^{\circ} \) and \( \phi^{2} = 45^{\circ} \), among others.
This model was first used in a Many-Worlds Monte Carlo minimization approach (a frequentist approach) to analyze the QWeak inelastic data, as implemented in \cite{androic2020parity} and \cite{JamesThesis}, and will also be used to reanalyze the data using the Bayesian analysis method, incorporating all inputs from Ref. \cite{androic2020parity}.

\begin{figure*}
\includegraphics[scale=0.35]{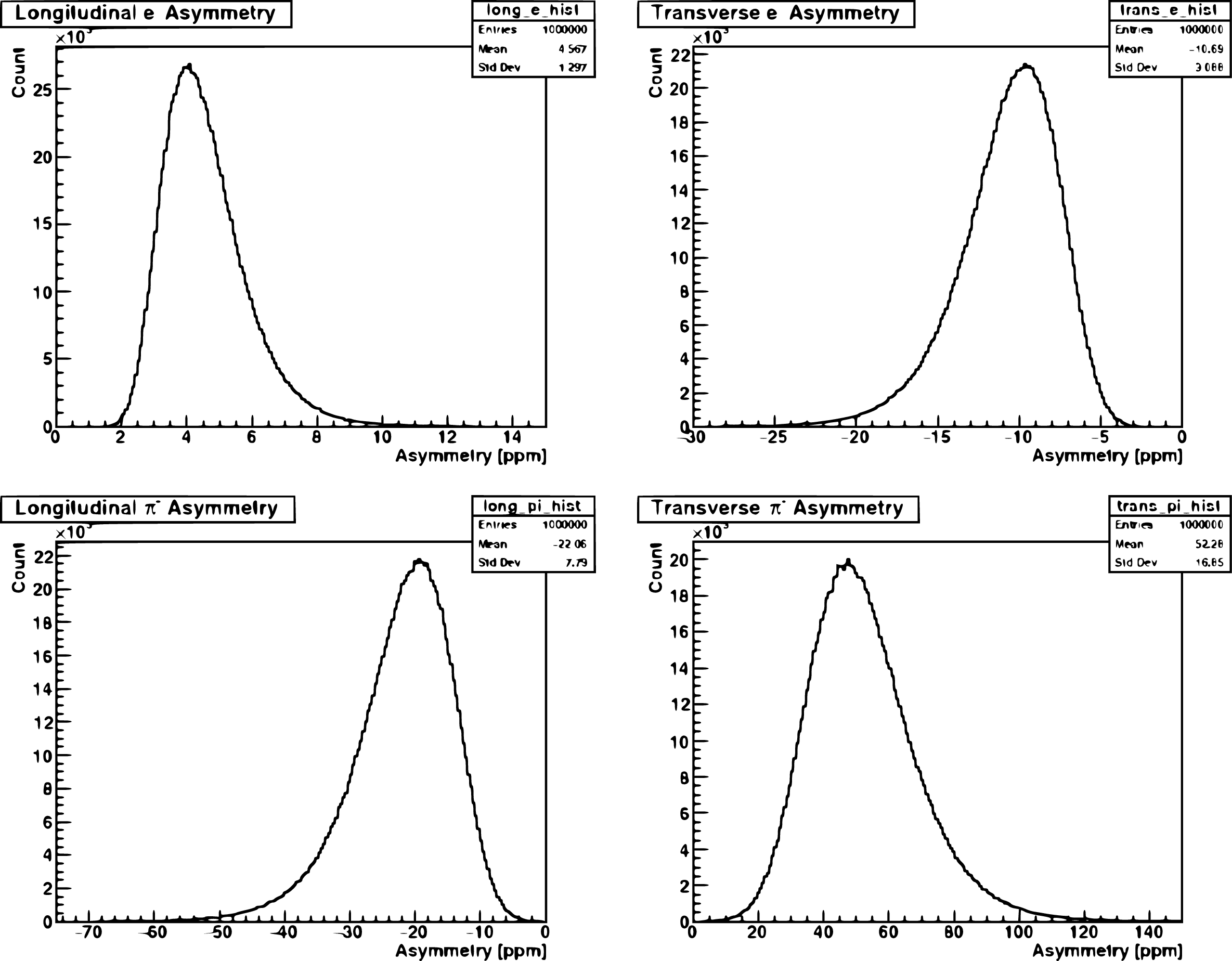}
\caption{\label{Figure 0}The distributions of the four asymmetries extracted through the Many-Worlds Monte Carlo minimization method, upper left: the asymmetry from electrons due to the longitudinal beam; upper right: the asymmetry from electrons due to the transverse beam; lower left: the asymmetry from pions due to the longitudinal beam; and lower right: the asymmetry from pions due to the transverse beam. It is important to note that there are differences in both sign and value between these histograms and the asymmetries presented in Table \ref{Table 1}, due to the beam polarization sign and value not being corrected during the initial analysis. The correct signs and values are those listed in Table \ref{Table 1} \cite{JamesThesis}.}
\end{figure*}

In the Many-Worlds Monte Carlo minimization approach, to extract the asymmetry components, \(A_e^L\), \(A_e^T\), \(A_{\pi}^L\), and \(A_{\pi}^T\), from the measured asymmetries in Equation \ref{MeasuredAsymmetriesQweak}, a value for each input quantity was randomly selected from a Gaussian distribution about their mean with widths equal to their uncertainties. These random values were then used to calculate the asymmetry in each MD \( A_{i,j}^{\text{calc}} \) and for each polarization configuration via the following equation \cite{JamesThesis}:
\begin{equation}
\begin{aligned}
A_{i,j}^{\text{calc}} = (1 - \tilde{f}_{\text{NB}}^i) \biggl[ &(1 - \tilde{f}_{\pi}^i)(A_e^L \cos \tilde{\theta}_P^j + A_e^T \sin \tilde{\theta}_P^j \sin \tilde{\phi}^i) \\
&+ \tilde{f}_{\pi}^{i}(A_{\pi}^L \cos \tilde{\theta}_P^j + A_{\pi}^T \sin \tilde{\theta}_P^j \sin \tilde{\phi}^i) \biggr],
\end{aligned}
\label{QweakMonteCarloModel}
\end{equation}
where a tilda over a quantity indicates a randomly selected value for that quantity. The function \(\delta\) \cite{JamesThesis}, where
\begin{equation}
    \delta_{\text{dof}}^2 = \sum \left( A_{i,j}^{\text{meas}} - A_{i,j}^{\text{calc}} \right)^2,
\label{Equation 5.20}
\end{equation}
was then minimized with respect to the asymmetries. This resulted in one possible set of values for each asymmetry, \( A_e^L \), \( A_e^T \), \( A_{\pi}^L \), and \( A_{\pi}^T \). The randomization and minimization process was repeated \( 10^6 \) times, giving \( 10^6 \) extracted values for each of the four asymmetries, which were used to shape the histograms shown in Fig.~\ref{Figure 0}. The root mean squared of the resulting distributions was taken as their uncertainties. The same data will be reanalyzed using the Bayesian approach, and the corresponding results will be presented and discussed in the next section.

\subsection{\label{MOLLER Experiment}MOLLER Experiment}
The MOLLER experiment aims to provide a precise measurement of the weak charge of the electron by measuring the parity-violating asymmetry in Møller scattering, which involves scattering longitudinally polarized electrons off unpolarized electrons, as detailed in \cite{benesch2014moller}. In the experiment, an 11 GeV electron beam with a current of 65 $\mu$A will be directed towards a 125 cm liquid hydrogen target. The electron beam's polarization, crucial for detecting the asymmetry, will be maintained at approximately 85\%. After the target, a spectrometer system composed of toroidal magnet assemblies and precision collimators with a seven-fold symmetry will separate Møller electrons from various backgrounds. It will direct them to a downstream detector plane. This system is optimized to maximize the signal-to-background ratio across the full azimuthal range and spans the polar angular range from \(5 \, \text{mrad} < \theta_{\text{lab}} < 21 \, \text{mrad}\). Møller scattering involves the scattering of identical particles, resulting in two electrons per event, emitted at opposite azimuthal angles and symmetrically around 90° in the center-of-mass frame. With its seven-fold azimuthal symmetry of blocked and unblocked sectors, the MOLLER apparatus is designed to accept either the forward- or backward-scattered electrons in each event.
The detector system includes 6 concentric rings of integrating detectors, the showermax detectors, and (encased in a lead absorber) a set of pion detectors. The integrating detectors, comprising concentric rings of fused silica with air-core light guides and photomultiplier tubes,  will measure both signal and background asymmetries. Notably, ring 5 will capture the Møller electron signal, while the other rings primarily record background interactions. The showermax detectors, designed to also intercept the Møller flux, are complemented by a lead absorber that reduces this flux before reaching the pion detectors. The pion detectors are acrylic Cherenkov detectors designed to measure pion dilutions and asymmetries.

As we transition from the technical specifications and operational details of the MOLLER experiment, it is instructive to compare its methodology with that of the QWeak data in terms of Bayesian analysis application. There are four distinct differences between the Qweak and MOLLER experiments when applying Bayesian analysis: experimental geometries, polarization angles, approaches to spin variation and data types. 
\\The experimental design differs significantly between the two; the QWeak data employed eight main detectors, with one acting as a pion detector, whereas the MOLLER experiment is planned to use a larger array of 84 main detector modules for ring 5 and 28 pion detector modules, necessitating an expanded Bayesian analysis framework. Note that in this paper, the analysis will focus on the main detector ring 5 and the pion detector. Eventually, we plan to expand this analysis to include all six rings of the main detectors, comprising 224 modules, the pion detector, and the showermax, each including 28 modules.

In terms of polarization angles, as mentioned before, the Qweak data involved a mixed polarization approach, with angles at \(-19.7^\circ \pm 1.9^\circ\) for the mixed data set and \(92.2^\circ \pm 1.9^\circ\) for the transverse data set. In contrast, in the MOLLER experiment, we assume that ideal longitudinal (\(0^\circ \pm 1^\circ\)) and transverse (\(90^\circ \pm 1^\circ\)) polarizations are applied for the longitudinal and transverse data sets, respectively.
The reason for maintaining a 1$^{\circ}$ uncertainty in the polarization angles within the MOLLER experiment arises from the need to suppress transverse polarization effects by averaging asymmetries over the full azimuth range. However, imperfect cancellation of these effects could lead to significant systematic errors. A transverse polarization component can introduce an azimuthal modulation of the measured asymmetry, potentially amplifying these systematic errors. To mitigate this, these polarization components will be continually monitored in the experimental hall and adjusted, by measuring the transverse scattering asymmetry with an accuracy below \(1^\circ\) within a few hours.

Regarding spin variations, the Qweak analysis implemented a simple model, assuming a predominantly horizontal transverse orientation \cite{androic2020parity}. In contrast, in the MOLLER experiment, more complex spin variation effects are integrated into its Bayesian framework. As a result of considering spin variation in the azimuthal plane (x-y plane), the spin angles in the longitudinal data-taking, when the polarization angle is set to \(0^\circ \pm 1^\circ \), span from \( -\pi \) to \( +\pi \), represented by \([-\pi, +\pi]\). In the transverse data-taking, when the polarization angle is set to \( 90^\circ \pm 1^\circ \), the spin angle is close to zero, as it is in the same plane as the polarization vector, represented by \( 0^\circ \pm 1^\circ\).

Lastly, the type of data used in the Bayesian analysis of the QWeak data comprised measured asymmetry values from the completed experiment. In contrast, for the MOLLER experiment, still in planning, we rely on synthetic or mock asymmetry values generated from simulations to model expected results and refine experimental parameters. Given the absence of empirical data from the MOLLER experiment, these synthesized data points are crucial for preparatory analyses.
It is assumed that the experiment will consist of two data-taking periods. The first one lasting 7,430 hours, represents 90\% of the total runtime, during which the polarization angle is set to a longitudinal configuration with an uncertainty of $1^{\circ}$. The second one lasting 825 hours, accounting for 10\% of the total runtime, has the polarization angle set to 90$^{\circ}\pm 1^{\circ}$  The parameters measured are referred to as mock asymmetry values, in contrast to the measured asymmetry values in QWeak for both the pion and main detectors.

Building upon these implementations, the asymmetry values for Møller electron and pion event generators, including longitudinal and transverse (vertical and horizontal) asymmetries, were simulated across various detectors (main detector ring 5 and pion detector). The contribution of these components to the final asymmetry values is modelled by 
\begin{equation}
\begin{aligned}
A_{i, j}^{\text{true}} &= (1 - f_{\pi}^{i}) \left[ A_{e}^L(i, j) \cos(\theta_{P}^j) + A_{e}^T(i, j) \sin(\theta_{P}^j) \right] \\
&\quad + f_{\pi}^{i} \left[ A_{\pi}^L(i, j) \cos(\theta_{P}^j) + A_{\pi}^T(i, j) \sin(\theta_{P}^j) \right] \\
A_{e}^T(i, j) &= A_{e}^{TV}(i, j) \sin(\phi_{P}^j) + A_{e}^{TH}(i, j) \cos(\phi_{P}^j) \\
A_{\pi}^T(i, j) &= A_{\pi}^{TV}(i, j) \sin(\phi_{P}^j) + A_{\pi}^{TH}(i, j) \cos(\phi_{P}^j)
\end{aligned}
\label{TrueAsymmetriesMOLLER}
\end{equation}
where \( A_{ij}^{\text{true}} \) represents the final expected asymmetry values from the simulations for each of the 84 modules in the main detector ring 5 and the 28 modules in the pion detector. The terms \( 1 - f_{\pi}^{i} \) and \( f_{\pi}^{i} \) denote the Møller and pion yield fractions, respectively, for each detector module \( i \). These fractions, which vary between the main and pion detectors, are calculated by dividing the number of generated photoelectrons from the pion generator by the total number of generated photoelectrons from both the pion and Møller generators at the detector modules. The longitudinal and transverse asymmetry values (vertical and horizontal) are denoted by \( A_{e}^L(i, j), A_{\pi}^{L}(i, j) \), \( A_{e}^{TV}(i, j), A_{e}^{TH}(i, j), A_{\pi}^{TV}(i, j), \) and \( A_{\pi}^{TH}(i, j) \), respectively. The angle \( \theta_{P}^j \) represents the polarization angle for dataset \( j \). The angle \( \phi_{P}^j \) is the angle in the azimuthal plane (x-y plane) from the x-axis to the projection of the spin vector onto this plane. The index \( i \) indicates the module number in the main or pion detector, while the index \( j \) corresponds to the dataset, whether longitudinal or transverse.

Each mock data element corresponds to the statistics of one hour of data-taking. Mock asymmetry values are generated using a normal distribution $\mathcal{N}(\mu,\sigma)$, where the derived true asymmetry values are the means $\mu$ and the measured uncertainties are the standard deviations $\sigma$. This approach introduces variation and realism into the mock data, \( A_{\text{mock}} \sim \mathcal{N}(A_{\text{true}}, \sigma) \). The measured uncertainties are calculated as the reciprocal of the square root of the product of their respective total rate of particles per detector module and the measurement time window (one hour). The total rate for each detector includes contributions from both Møller electron and pion generators. 
After generating mock asymmetry values, we can input them into 
\begin{equation}
\begin{aligned}
A_{i, j}^{\text{mock}} &= (1 - f_{\pi}^{i})\bigl[A_{i, j}A_{e}^L \cos(\theta_{P}^j) + C_{i, j} A_{e}^T \sin(\theta_{P}^j)\bigr] \\
&\quad + f_{\pi}^{i} [B_{i, j} A_{\pi}^{L} \cos(\theta_{P}^j) + D_{i, j} A_{\pi}^{T} \sin(\theta_{P}^j)] \\
A_{i, j} &= N_{A_e^{L}(i, j)}\\
B_{i, j} &= N_{A_{\pi}^{L}(i, j)}\\
C_{i, j} &= N_{A_e^{TV}(i, j)} \sin(\phi_{P}^j) + N_{A_e^{TH}(i, j)} \cos(\phi_{P}^j) \\
D_{i, j} &= N_{A_{\pi}^{TV}(i, j)} \sin(\phi_{P}^j) + N_{A_{\pi}^{TH}(i, j)} \cos(\phi_{P}^j)
\end{aligned}
\label{MockAsymmetriesMOLLER}
\end{equation}
where \(A_{i, j}^{\text{mock}}\) represents the generated mock asymmetry values for the pion and main detector modules. \(A_e^L\), \(A_e^T\), \(A_{\pi}^L\), and \(A_{\pi}^T\) represent the Møller and pion asymmetry components that need to be extracted. The terms \(A_{ij}\), \(B_{ij}\), \(C_{ij}\), and \(D_{ij}\) are referred to as kinematic coefficients, which best replicate the seven-fold symmetry nature of the experiment and the existence of identical-particle scatterings discussed before (in contrast to the case of the QWeak data). \(N_{A_{e}^L}\), \(N_{A_{\pi}^L}\), \(N_{A_e^{TV}}\), \(N_{A_e^{TH}}\), \(N_{A_{\pi}^{TV}}\), and \(N_{A_{\pi}^{TH}}\) are normalized asymmetry values obtained by normalizing \(A_e^{L}\), \(A_{\pi}^{L}\), \(A_e^{TV}\), \(A_e^{TH}\), \(A_{\pi}^{TV}\), and \(A_{\pi}^{TH}\) in Eq.~\ref{TrueAsymmetriesMOLLER}. The key point is that the normalizing factors for obtaining the normalized asymmetry values should be independent of the experiment’s kinematics. To calculate these normalizing factors, asymmetries are simulated within the acceptance range of the MOLLER experiment, and the maximum values obtained are considered normalizing factors. Other parameters align with those in Eq.~\ref{TrueAsymmetriesMOLLER}. 
This model will be used here to analyze the simulated MOLLER experiment, incorporating all the assumed inputs.
\section{\label{Validation of the Results}Validation of the Results}
This section illustrates the results of implementing Bayesian analysis using the inputs and models from the QWeak and MOLLER experiments. For QWeak, validation is achieved by comparing these results with those obtained from the Many-Worlds Monte Carlo minimization approach \cite{androic2020parity,JamesThesis}. For MOLLER, validation is conducted by testing agreement with the input asymmetries in the simulations.
\subsection{\label{Reanalysis of QWeak Experiment}Reanalysis of the QWeak Inelastic Experiment}
\begin{figure*}
\includegraphics[scale=0.64]{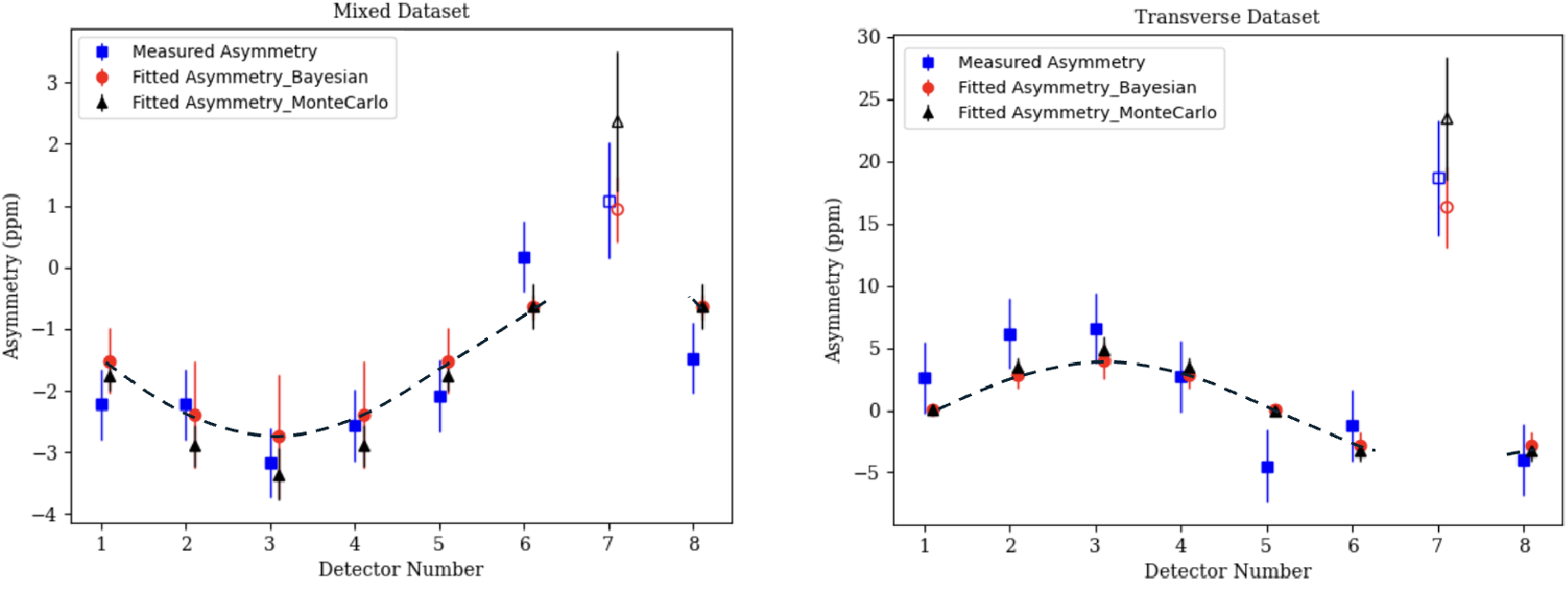}
\caption{\label{Figure 1}Comparative analysis of the measured inelastic QWeak and fitted asymmetry values across detectors for two data sets. On the left, the mixed data set is shown with measured asymmetry depicted as blue squares, compared with values derived through Bayesian inference (red circles) and Monte Carlo simulations (black triangles). On the right, the transverse data set is similarly represented. The data spans eight detectors in each plot, with error bars indicating the uncertainty for each data point. MD7 is highlighted with hollow symbols to indicate its distinct characteristics from the other MDs. Dashed lines in both data sets show the fitted sine wave in the Bayesian analysis, which is different from those obtained using the Monte Carlo method  \cite{androic2020parity}.}
\end{figure*}

Incorporating all inputs from the QWeak inelastic data, \(A^{\text{meas}}\), \(f_{\pi}\), \(f_{\text{NB}}\), \(\theta\), and \(\phi\) (with \(y\) as defined in Eq.~\ref{BayesRuleOriginal}), along with the model described by Eq.~\ref{MeasuredAsymmetriesQweak}, the asymmetry values for Møller electrons and pions, \(A_e^L\), \(A_e^T\), \(A_{\pi}^L\), and \(A_{\pi}^T\) (with \(\theta\) as defined in Eq.~\ref{BayesRuleOriginal}), have been calculated using the Many-Worlds Monte Carlo Minimization method \cite{androic2020parity,JamesThesis} and are now recalculated using Bayesian analysis.
\\The results for the extracted asymmetry values from both analysis methods are presented in parts per million (ppm) in Table \ref{Table 1}.
\begin{table}[b]
\caption{\label{Table 1}Comparison of asymmetry values and uncertainties for QWeak inelastic data: Many-Worlds Monte Carlo Minimization Method (MC) versus Bayesian analysis (B).}
\begin{ruledtabular}
\begin{tabular}{ccccc}
\textrm{Method}&
\textrm{\( A_e^L ({\rm ppm}) \)}&
\textrm{\( A_e^T ({\rm ppm})\)}&
\textrm{\( A_{\pi}^L ({\rm ppm}) \)}&
\textrm{\( A_{\pi}^T ({\rm ppm})\)}\\
\colrule
MC & -5.25 $\pm$ 1.49 & 12.3 $\pm$ 3.6 & 25.4 $\pm $9.0 & -60.1 $\pm$ 19.3\\
B & -4.9$ \pm$ 0.7 & 12.0$ \pm$ 2.0 & 22.8 $\pm$ 5.9 & -55.9 $\pm$ 14.5\\
\end{tabular}
\end{ruledtabular}
\end{table}
The comparison shows that the Monte Carlo minimization method consistently resulted in larger absolute values and uncertainties across all measured asymmetries compared to the Bayesian analysis. To evaluate the performance of both analysis techniques, asymmetry values derived using both approaches were substituted into Eq.~\ref{MeasuredAsymmetriesQweak}. The resulting asymmetry values, referred to as fitted asymmetries, along with their associated uncertainties, are presented for the two data sets across detectors (MDs) in the plots of Fig.~\ref{Figure 1}. 

In both data sets, the Bayesian analysis method (represented by red circles) demonstrates a closer fit to the measured asymmetry (depicted by blue squares) compared to the Monte Carlo minimization method (shown as black triangles), with a calculated chi-squared per degree of freedom value of 1.2 for the Bayesian method compared to 1.3 for the Monte Carlo method. The error bars, representing uncertainties in the measured asymmetries, are consistently smaller in the Bayesian approach compared to those from the Monte Carlo minimization method. Furthermore, in contrast to the Monte Carlo method, the error bars associated with the Bayesian method show less variation across different detector numbers, indicating a more stable estimation process.

This is explained by the mechanisms of both analysis methods. As outlined in the methodology, Bayes' theorem involves an iterative process in which the posterior distribution from each iteration serves as the prior for the next, incorporating new data sets. This approach ensures that, in the absence of informative prior knowledge, the Bayesian analysis remains data-driven. Additionally, uncertainty is quantified by the width of the posterior distribution in the final step, which is shaped by the data and leads to more precise estimations. 

In contrast, the Monte Carlo minimization method also explained in the methodology section, derives asymmetry components from measured data by randomly sampling input values from Gaussian distributions based on their means and uncertainties. These values are used to calculate asymmetries, and the method minimizes the squared deviations between measured and calculated asymmetries to estimate the unknown components. This process generates distributions for each component, as shown in Figure \ref{Figure 0}, and the RMS of these distributions is used to determine the uncertainty for each component, resulting in a tail in the histogram and larger uncertainty estimates due to the broader spread of possible values. 

The results of this comparison suggest that assuming Gaussian distributions for inputs that are likely non-Gaussian, such as \(f_{\pi}\) (which is strictly  bound in the interval $[0,1]$), is problematic in the Monte Carlo minimization method. In a test study, when we reduced the uncertainty on \(f_{\pi}\) by one order of magnitude, the tails of the distributions in Figure \ref{Figure 0} disappeared, and the uncertainties became closer to those obtained from the Bayesian analysis. This study confirmed that the assumption of Gaussianity for all inputs is imprecise. Furthermore, the method used to calculate uncertainty was found to be inaccurate. The RMS of the distributions of squared deviations between measured and calculated asymmetries is not a precise way to quantify uncertainty. Addressing these issues should ultimately lead to consistent results from both methods.

We note that, for each of the four extracted asymmetries, the central values of the new Bayesian results are in agreement within uncertainties with the originally-reported Monte Carlo method results, and so the physics conclusions reported in Ref.~\cite{androic2020parity} are unchanged by this new analysis.
\subsection{\label{Analysis of MOLLER Experiment}Analysis of the MOLLER Experiment}
Incorporating the simulations and assumptions from the MOLLER experiment, the input data include \(A^{\text{mock}}\), \( f^i_{\pi} \), \(\theta_{P}^j\), \(\phi_{P}^j\), \(A_{ij}\), \(B_{ij}\), \(C_{ij}\), and \(D_{ij}\) (i.e. these are the \(y\) in Eq.~\ref{BayesRuleOriginal}). The parameters to be extracted are the asymmetry values \(A_e^L\), \(A_e^T\), \(A_{\pi}^L\), and \(A_{\pi}^T\) (i.e. these are the \(\theta\) in Eq.~\ref{BayesRuleOriginal}). The model, specified by Eq.~\ref{MockAsymmetriesMOLLER}, is applied separately to the pion and main detectors for both longitudinal and transverse measurements. Consequently, the asymmetry values are extracted using four equations: two for the pion detector and two for the main detector, covering both longitudinal and transverse measurements. The results of the extracted asymmetry values and their associated uncertainties in ppb are summarized in Table \ref{Table 2} and compared with the inputs of the analysis.
\begin{table}
\caption{\label{Table 2}Comparison of asymmetry values and uncertainties: Inputs versus Outputs for MOLLER simulated data.}
\begin{ruledtabular}
\begin{tabular}{ccccc}
 & 
\textrm{\( A_e^L \) ({\rm ppb})} & 
\textrm{\( A_e^T \) ({\rm ppb})} & 
\textrm{\( A_{\pi}^L \) ({\rm ppb})} & 
\textrm{\( A_{\pi}^T \) ({\rm ppb})} \\
\colrule
Inputs & -28.00 & 13845 & 28400 & -53667 \\
Outputs & -28.40 & 13823.0 & 28486 & -53762 \\
 & $\pm$ 0.51 & $\pm$ 9.6 & $\pm$ 75 & $\pm$ 190\\
\end{tabular}
\end{ruledtabular}
\end{table}
For \(A_e^L\), the input value of -28.00 ppm closely matches the output value of -28.40 ppm with an uncertainty of ±0.51 ppm, indicating strong agreement. However, for \(A_e^T\), the input value is 13,845 ppm, and the corresponding output is 13,823.0 ppm with an uncertainty of ±9.6 ppm, showing a slight discrepancy outside the uncertainty range. The \(A_{\pi}^L\) input of 28,400 ppm is slightly exceeded by the output of 28,486 ppm with an uncertainty of ±75 ppm, and the \(A_{\pi}^T\) input of -53,667 ppm is well-approximated by the output of -53,762 ppm, with a larger uncertainty of ±190 ppm. Overall, the close alignment between most inputs and outputs within the uncertainty ranges validates the effectiveness of the Bayesian model in accurately reproducing the asymmetry values, confirming the reliability and precision of the analysis.

\begin{figure}
\includegraphics[scale=0.56]{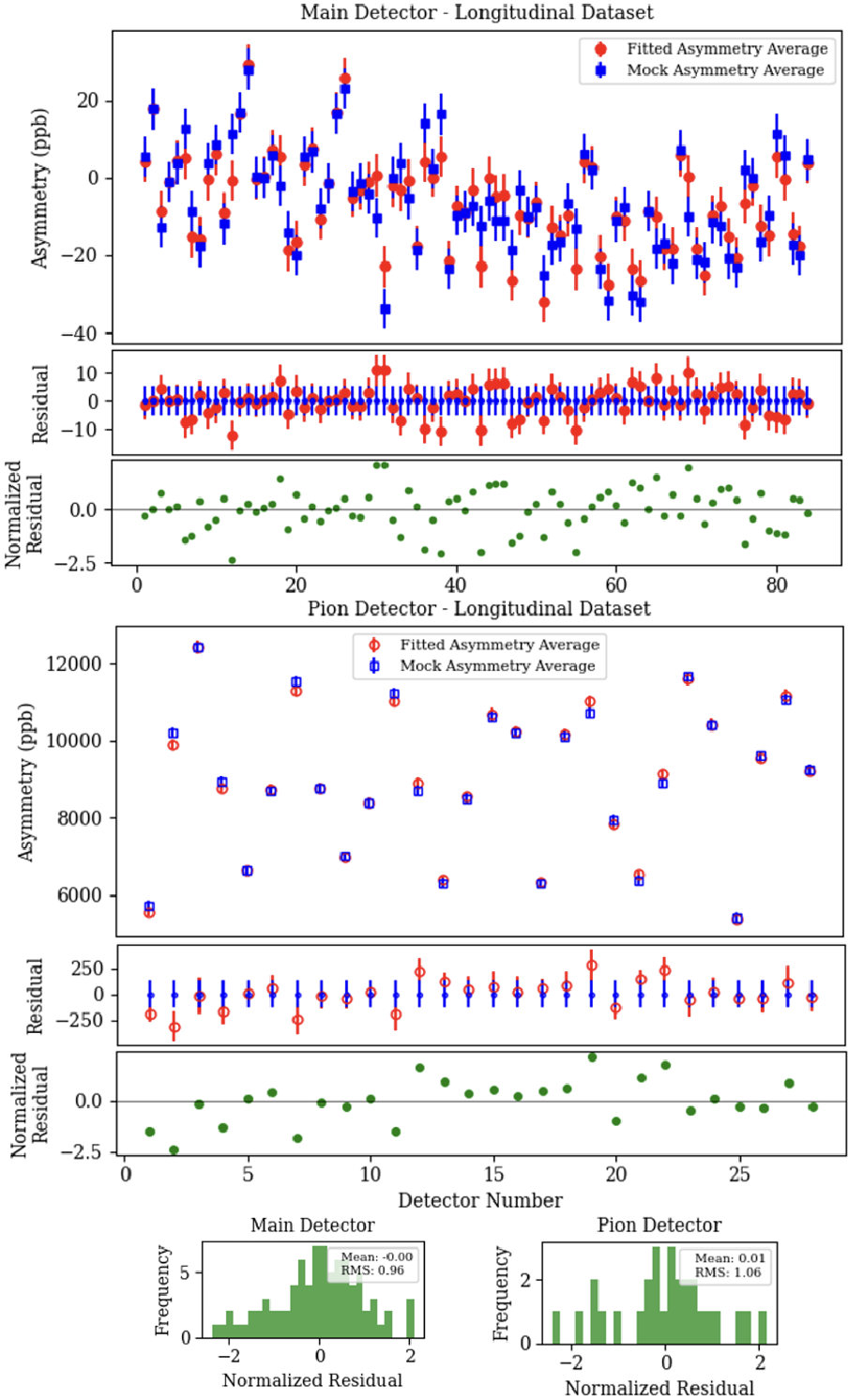}
\caption{\label{Figure 2}Comparison of average fitted and mock asymmetry values in the MOLLER experiment's main (top plot) and pion (bottom plot) detectors in the longitudinal data set. Blue
squares represent mock values, and red circles indicate fitted
values, with error bars for uncertainty. Middle plots highlight residuals to underscore small error bars. Bottom plots display normalized residuals and histograms, evaluating fit quality through means and RMS values. The pion detector is highlighted with hollow symbols to indicate its distinct characteristics from the main detector.}
\end{figure}

\begin{figure}
\includegraphics[scale=0.615]{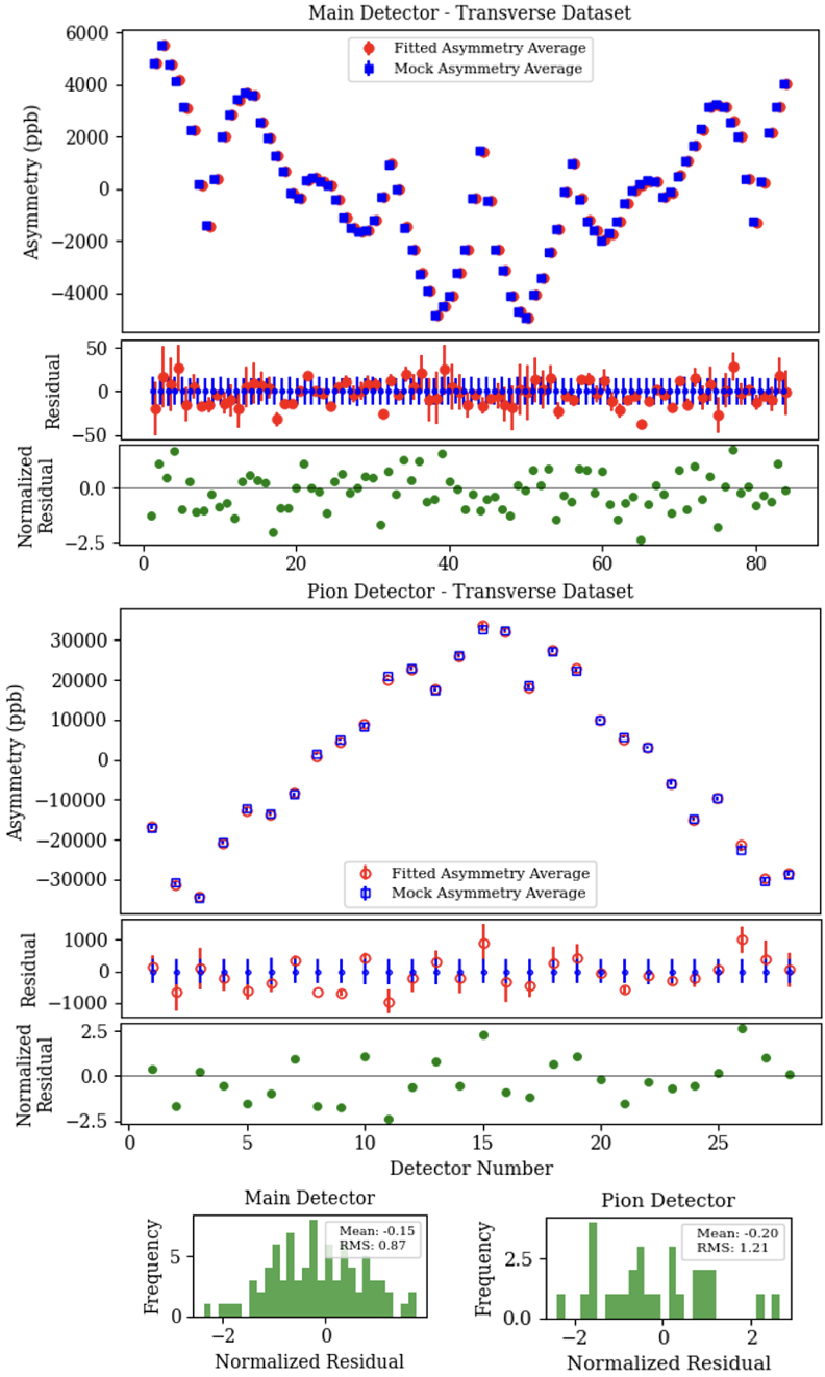}
\caption{\label{Figure 3}Comparison of average fitted and mock asymmetry values in the MOLLER experiment's main (top plot) and pion (bottom plot) detectors in the transverse data set. Blue squares represent mock values, and red circles indicate fitted values, with error bars for uncertainty. Middle plots highlight residuals to underscore small error bars. Bottom plots display normalized residuals and histograms, evaluating fit quality through means and RMS values. The pion detector is highlighted with hollow symbols to indicate its distinct characteristics from the main detector.}
\end{figure}

Another way to verify the results is by substituting the asymmetry values along with other parameters into the right-hand side of Eq.~\ref{MockAsymmetriesMOLLER}, computing and deriving another set of asymmetry values, referred to as the fitted asymmetry values, similar to the case of Qweak. 
Figures \ref{Figure 2} and \ref{Figure 3} provide a comparative analysis of fitted versus mock asymmetry for the main and pion detectors in both the longitudinal and transverse data sets. In the top plots, the blue squares represent the average of mock asymmetry values, while the red circles denote the average of fitted asymmetry values obtained through Bayesian analysis. Due to the large scale of the asymmetry values, the error bars are too small to be visible in some of these plots. To address this, the middle plots subtract the mock asymmetry average from both the mock and fitted asymmetry values, effectively re-centering the data around zero. This re-centering provides a clearer visualization of the error bars, illustrating the measurement precision. In these plots, the blue dots represent the residual mock asymmetry values, which are zero as they are subtracted from themselves. The red circles display the residual fitted asymmetry values, with the error bars indicating the uncertainties. The bottom plots present histograms of the normalized residuals, defined as the difference between the fitted and mock asymmetry values, normalized by the uncertainty in the mock asymmetry values. These histograms help assess the normality and spread of these residuals. 

In the longitudinal data set, the main detector shows the residuals centred around a mean of 0.00 with an RMS of 0.96. In the case of the pion detector, the residuals have a mean of -0.01 and an RMS of 1.06. In both cases, the residuals are distributed close to zero, indicating a good fit. For the transverse data set, the main detector residuals have a mean of -0.15 and an RMS of 0.87, while the pion detector residuals have a mean of -0.20 and an RMS of 1.21. These larger discrepancies likely arise from the smaller set of transverse measurements, a tenth of the longitudinal measurements, resulting in decreased statistical power. Similar to MD7 in Fig.~\ref{Figure 1}, the pion detector is highlighted with hollow symbols to indicate its distinct characteristics from the main detector.
These plots are used to validate the Bayesian framework and ensure the model's robustness in predicting asymmetry values before having actual experimental data to apply it to. As shown, the longitudinal asymmetry at the pion detector varies rapidly between the 7-fold open and closed sectors, while the transverse asymmetry varies predominantly as a cosine function of the azimuthal angle. The longitudinal asymmetry at the main detector is a constant value, while the variation in the transverse asymmetry arises from the mixture of two electrons moving in opposite azimuthal directions, with different probabilities of acceptance or rejection for each electron in different detector modules.
\section{\label{Discussion}Discussion}
Understanding the correlation between parameters is essential for revealing the complex relationships that govern the data. By quantifying how different variables co-vary, we can gain deeper insights into their interdependencies and improve the precision of parameter estimates. To achieve this, the covariance matrix provides a valuable tool for evaluating these correlations. The covariance matrix is calculated as follows:
\begin{equation}
\text{Cov}(X, Y) = \frac{1}{n-1} \sum_{i=1}^{n} (X_i - \bar{X})(Y_i - \bar{Y})
\end{equation}
where \(\text{Cov}(X, Y)\) is the covariance between two variables \(X\) and \(Y\), \(X_i\) and \(Y_i\) are individual observations from variables \(X\) and \(Y\) respectively, \(\bar{X}\) and \(\bar{Y}\) are the means of \(X\) and \(Y\), and \(n\) is the number of observations.
For the QWeak and MOLLER experiments studied here, the covariance matrix is a \( 4 \times 4 \) square matrix representing the relationships between determined longitudinal and transverse asymmetries of electrons and pions. In this matrix, the off-diagonal elements reflect the covariance between the asymmetry components. For instance, \(\text{Cov}(A_e^L, A_e^T)\) quantifies the degree to which the longitudinal and transverse electron asymmetries co-vary, providing insights into how these variables influence each other.
The correlation matrix normalizes the covariances by the standard deviations of the variables involved as follows:
\begin{equation}
\text{Corr}(X, Y) = \frac{\text{Cov}(X, Y)}{\text{Cov}(X, X) \text{Cov}(Y, Y)}, \quad \text{if} \ \sigma_X \sigma_Y > 0.
\end{equation}
This equation provides a scaled representation highlighting the strength and direction of the linear relationships between variables independent of their units.

\begin{figure}[h!]
\includegraphics[scale=0.29]{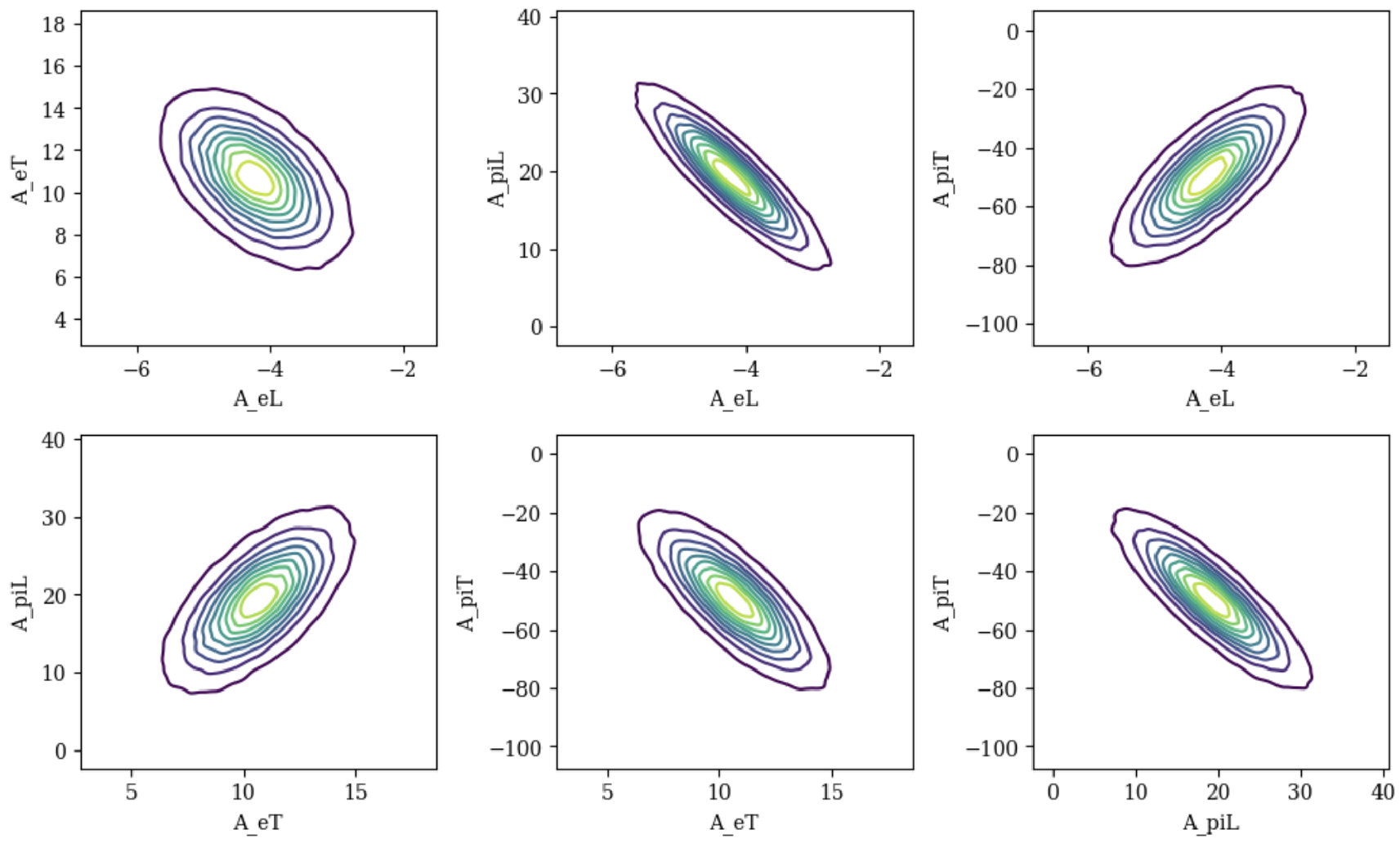}
\caption{\label{Figure 4}Correlation contours between the parameters \(A_e^L\), \(A_e^T\), \(A_{\pi}^L\), and \(A_{\pi}^T\) in the QWeak experiment, derived using Bayesian analysis, with decile lines marking the density distribution from 10\% to 90\%. Each plot illustrates the correlation between two parameters, showcasing the density and direction of their relationships.}
\end{figure}

Figure \ref{Figure 4} visualizes these correlations for longitudinal and transverse electron and pion asymmetries, derived from the Monte Carlo minimization method and Bayesian analysis in the QWeak experiment.
Each contour plot is a two-dimensional kernel density estimation that shows how the values of one parameter vary with another. The decile lines, marking the density distribution from 10\% to 90\%, serve to highlight areas where data points are most densely concentrated, which are indicative of strong correlations.
It is evident that \(A_e^L\) is negatively correlated with \(A_e^T\), similar to the negative correlation observed between \(A_{\pi}^L\) and \(A_{\pi}^T\). As previously discussed, pure longitudinal polarization results in pure longitudinal asymmetry and deviations from this ideal state give rise to transverse asymmetry, thereby causing these negative correlations. Additionally, there is a strong negative correlation between \(A_e^L\) and \(A_{\pi}^L\), as well as between \(A_e^T\) and \(A_{\pi}^T\). In contrast, \(A_e^T\) is positively correlated with \(A_{\pi}^L\), and \(A_e^L\) is positively correlated with \(A_{\pi}^T\). 

\begin{figure}[h!]
\includegraphics[scale=0.34]{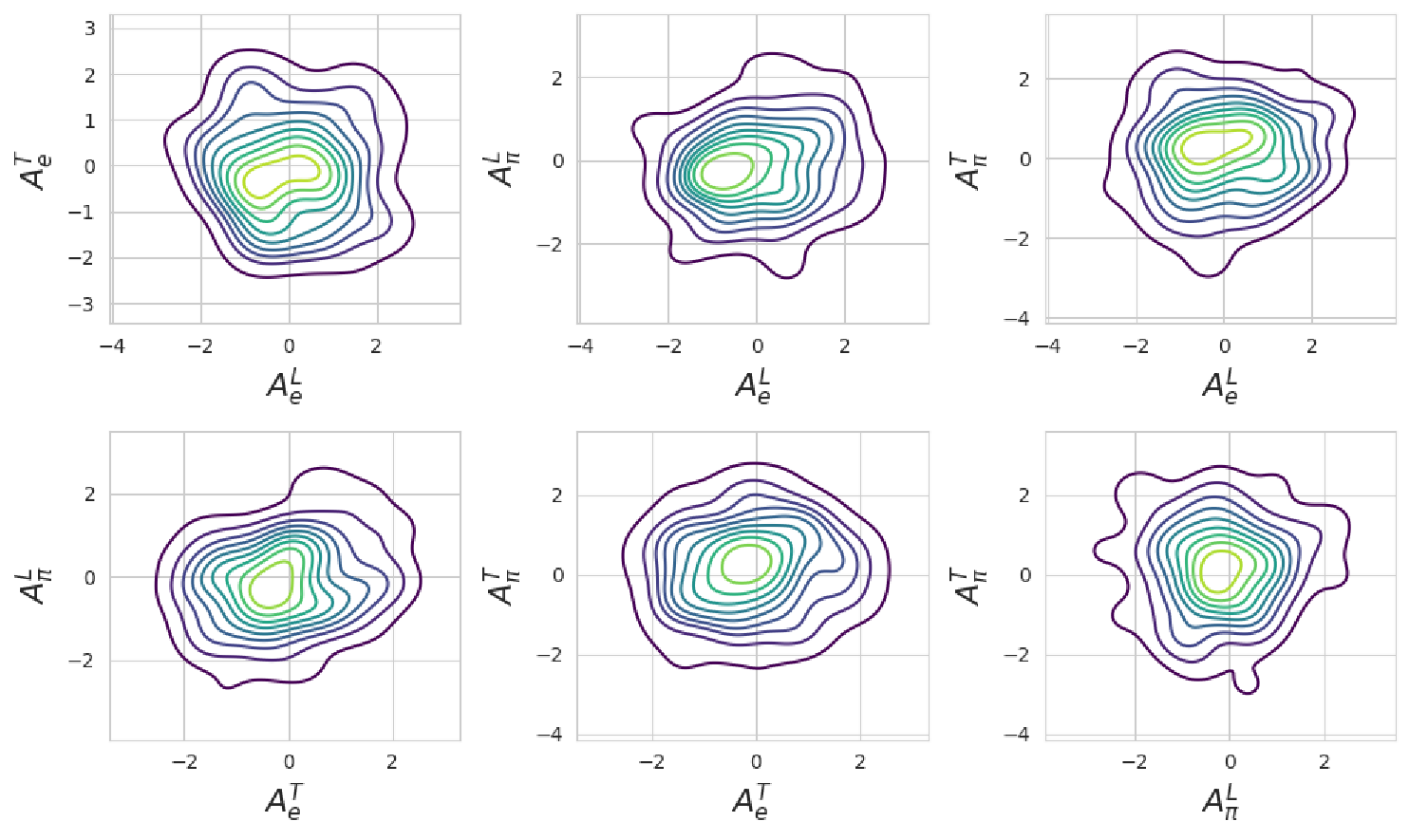}
\caption{\label{Figure 5}Correlation contours between the parameters \(A_e^L\), \(A_e^T\), \(A_{\pi}^L\), and \(A_{\pi}^T\) in the MOLLER experiment, derived using Bayesian analysis, with decile lines marking the density distribution from 10\% to 90\%. Each plot illustrates the correlation between two parameters, showcasing the density and direction of their relationships.}
\end{figure}

Determining the correlations between asymmetry components in the MOLLER experiment is the final aspect of our discussion. As seen in Fig.~\ref{Figure 5}, there is no correlation between the estimated parameters \(A_e^L\), \(A_e^T\), \(A_{\pi}^L\), and \(A_{\pi}^T\). This absence of correlation demonstrates how the MOLLER setup and kinematics successfully separate the main detector signals and the pion detector signals and thus will provide independent, uncontrolled access to both. 

\section{\label{Conclusion}Conclusion}
In this study, the reanalysis of the QWeak inelastic data using the Bayesian method, compared to the previously employed Many-Worlds Monte Carlo Minimization approach, provided insights into the precision and reliability of both techniques. The Bayesian analysis produced smaller uncertainties and more stable estimations, offering a closer fit to the measured asymmetry values. In contrast, the Monte Carlo method resulted in larger uncertainties and greater variations across detectors, likely due to the incorrect assumption of Gaussian distributions for non-Gaussian inputs and the approach to quantify the uncertainties. Despite these differences, both methods yielded central values consistent within their respective uncertainties, ensuring that the physics conclusions of the original study remained unchanged. The analysis of the MOLLER experiment using Bayesian inference demonstrated an agreement between the simulated input asymmetry values and the extracted output values within their respective uncertainty ranges. The comparative analysis of fitted versus mock asymmetry values further validated the Bayesian approach, with the residuals centered around zero and exhibiting small RMS values. Furthermore, the correlation analysis in both experiments revealed insights into the interdependencies of asymmetry components, with QWeak showing expected correlations between longitudinal and transverse asymmetries, while the MOLLER experiment's setup effectively separated detector signals, resulting in no correlations between asymmetry parameters.
The promising outcomes from the QWeak and MOLLER experiments establish a foundation for applying Bayesian analysis to upcoming high-precision parity-violation experiments such as P2 and SoLID. This method is anticipated to be equally effective in addressing the unique challenges presented by these experiments.

\begin{acknowledgments}
This work was supported by the National Science Foundation (grants PHY-2012738 and PHY-2412825) and the Natural Sciences and Engineering Research Council of Canada
(grant SAPPJ-2022-00019). We are grateful to our colleagues in the QWeak and MOLLER collaborations, and in particular to Z.S. Demiroglu for help with the MOLLER simulation framework and J.F. Dowd for help with the Qweak data. 
\end{acknowledgments}

\nocite{*}
\bibliography{Enhancing_Precision_in_PVES_Experiments}% Produces the bibliography via BibTeX.

\end{document}